\documentstyle[12pt,aps,psfig]{revtex}
\begin{document}
\newcommand{\be} {\begin{equation}}
\newcommand{\ee} {\end{equation}}
\newcommand{\bea} {\begin{eqnarray}}
\newcommand{\eea} {\end{eqnarray}}
\newcommand{\ba} {\begin{array}}
\newcommand{\ea} {\end{array}}
\newcommand{\nn} {\nonumber}
\newcommand{\noi} {\noindent}
\title{\Large
  Distribution of Avalanche Sizes in the Hysteretic Response 
  of Random Field Ising Model on a Bethe Lattice at Zero Temperature }
\author{\large 
Sanjib Sabhapandit$^\S$\footnote[1]{E-mail address:
  sanjib@theory.tifr.res.in} 
\and  Prabodh Shukla$^\P$\footnote[2]{E-mail address:
  shukla@dte.vsnl.net.in} 
\and  Deepak Dhar$^\S$\footnote[3]{E-mail address:
  ddhar@theory.tifr.res.in } 
}
\address{$^\S$Theoretical Physics Group, Tata Institute of 
Fundamental Research, \\
            Homi Bhabha Road, Mumbai-400005, India.\\
$^\P$Physics Department, North Eastern Hill University, \\
                        Shillong-793 022, India.
}
\maketitle


\begin{abstract}

We consider the zero-temperature single-spin-flip dynamics of the
random-field Ising model on a Bethe lattice in the presence of an external
field $h$. We derive the exact self-consistent equations to determine the
distribution $Prob(s)$ of avalanche sizes $s$, as the external field
increases from $-\infty$ to $\infty$. We solve these equations explicitly
for a rectangular distribution of the random fields for a linear chain and
the Bethe lattice of coordination number $z=3$, and show that in these
cases, $Prob(s)$ decreases exponentially with $s$ for large $s$ for all
$h$ on the hysteresis loop. We find that for $z \ge 4$ and for small
disorder, the magnetization shows a first order discontinuity for several
continuous and unimodal distributions of the random fields.  The avalanche
distribution $Prob(s)$ varies as $s^{-3/2}$ for large $s$ near the
discontinuity.

\end{abstract}


  
\section{Introduction}
\label{intro}

Analytical treatment of problems having quenched disorder is usually
difficult. There are few models having nontrivial quenched disorder that
can be solved exactly.  In this paper, we obtain exact results for the
non-equilibrium properties of the random-field Ising model (RFIM) on the
Bethe lattice. We consider the single-spin-flip Glauber dynamics of the
system at zero temperature, as the external magnetic field is slowly
varied from $-\infty$ to $+\infty$. As the field increases, the
magnetization increases as groups of spins flip up together.  This model
has been proposed as a model of the Barkhausen noise by Sethna {\it et
al}~\cite{sethna} (see also ~\cite{PDS}). In this paper, we set up the
exact self-consistent equations satisfied by the generating function of
the distribution of avalanche sizes, and analyze these to determine the
behavior of the avalanche distribution function on the Bethe lattice.

The study of the equilibrium properties of the RFIM has been an important
problem in statistical physics for a long time.  In 1975, Imry and Ma
\cite{imryma}, showed that arbitrarily weak disorder destroys long-ranged
ferromagnetic order in dimensions $ d < 2$. The persistence of
ferromagnetism in $ d=2$ was a matter of a long controversy, but has now
been established \cite{imbrie}.  A recent review of earlier work on this
model may be found in \cite{natterman}. As far as an exact calculation of
thermodynamic quantities is concerned, there are only a few results. For
example, Bruinsma studied the RFIM on a Bethe lattice in the absence
of an external field and for a bivariate random field
distribution~\cite{bruinsma}. There are no known exact results for the
average free energy or magnetization, for a continuous distribution of
random field, even at zero temperature and in zero applied field.

\begin{figure}
\begin{center}
\leavevmode
\psfig{figure=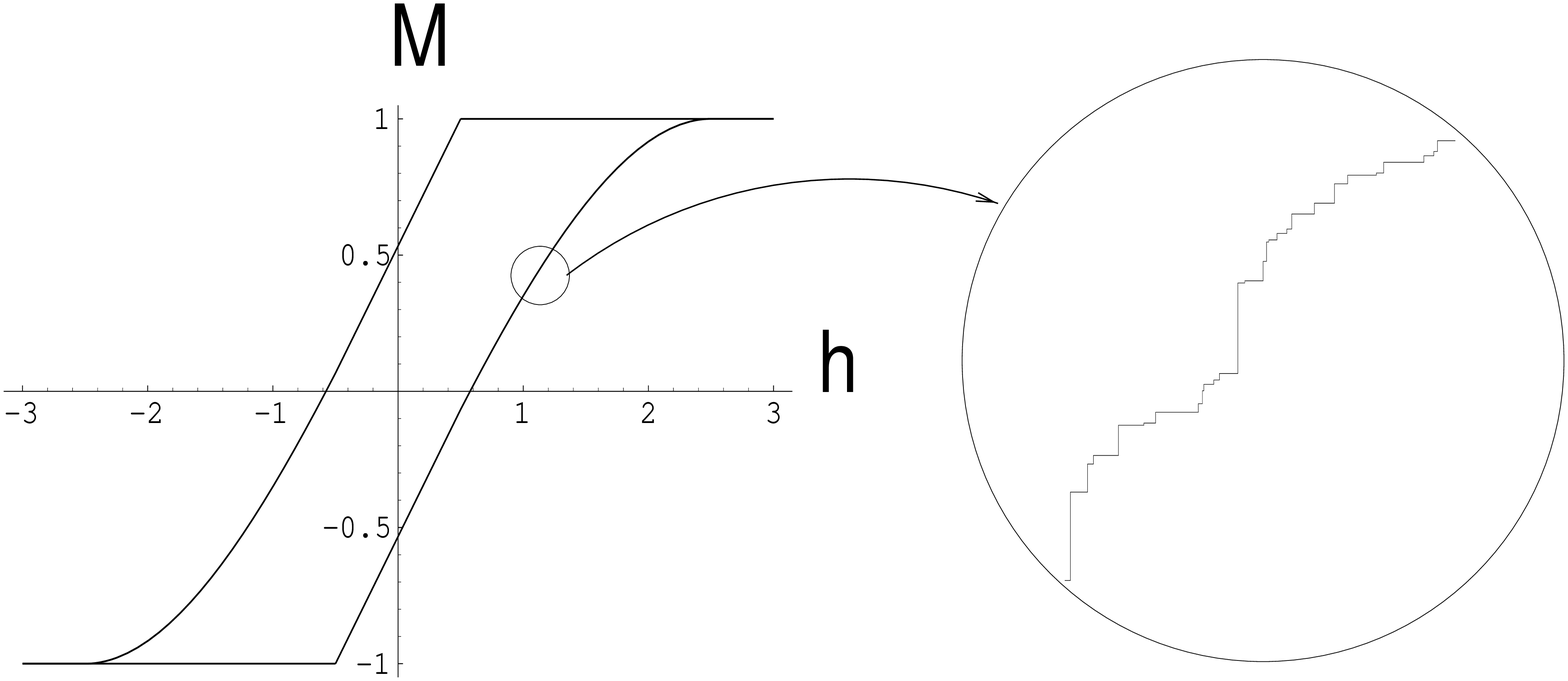,width=12cm}
\caption{The Hysteresis loop of magnetization $M$ versus the external
field $h$ for RFIM. The zoomed figure shows the small jumps in
magnetization that give rise to the Barkhausen noise}
\label{barkhausen}
\end{center}
\end{figure}

The non-equilibrium properties of the RFIM has attracted a lot of
interest lately, arising from the observation by Sethna {\it et al}
~\cite{sethna} 
that its zero-temperature dynamics provides a simple model for the
Barkhausen noise and return point memory.  Barkhausen noise is the high
frequency noise generated due to the small jumps in magnetization observed
when ferromagnets are placed in oscillating magnetic fields
[Fig.~\ref{barkhausen}]. 
Understanding and reduction of this noise is important for the design of
many electronic devices~\cite{sipahi}. Experimentally it is
observed~\cite{McClure,brien,Urbach,cote} that the the increase of the
magnetization occurs in bursts that span over two decades of size and the
distribution of burst (avalanche) sizes seems to follow a power law 
in this range. Similar avalanche-like relaxational
events are also observed in other systems, for example, the stress-induced
martensitic growth in some alloys~\cite{carrillo}. This power-law
tail in the event-size distribution was interpreted by Cote and
Meisel~\cite{cote} as an  
example of {\it self-organized criticality}.  But Percovi\'{c} {\it et al}
~\cite{PDS} have argued that large bursts are exponentially rare, and the
approximate power-law tail of the observed distribution comes from
crossover effects due to nearness of a critical point. Recently Tadi\'{c}
\cite{tadic} has presented some evidence from numerical simulations that
the exponents for avalanche distribution can vary continuously with
disorder. Our results about the behavior of the avalanche distribution
function also relate to this question whether any fine-tuning of
parameters is required to see power-law tails in the avalanche
distribution in the RFIM, and if the exponents can be varied
continuously with disorder.

The advantage of working on the Bethe lattice is that the usual BBGKY
hierarchy of equations for correlation functions closes, and one can hope
to set up exact self-consistent equations for the correlation functions.
The fact that Bethe's self-consistent approximation becomes exact on the
Bethe lattice is useful as it ensures that the approximation will not
violate any general theorems, e.g. the convexity of thermodynamic
functions, sum rules. In the presence of disorder, in spite of the
closure of the BBGKY hierarchy, the Bethe approximation is still very
difficult, as the self-consistent equations become functional equations
for the probability distribution of the effective field. These are not
easy to solve, and available analytical results in this direction are
mostly restricted to one dimension \cite{1dsolns}, or to models with
infinite-ranged interactions \cite{SK}. On the Bethe lattice, for
short-ranged interactions with quenched disorder, e.g. in the prototypical
case of the $\pm J$ random-exchange Ising model, the average free energy
is trivially determined in the high temperature phase, but not in the
low-temperature phase. It has not been possible so far to determine even
the ground-state energy exactly despite several attempts
\cite{sherrington.katsura}.

Calculation of time-dependent or non-equilibrium properties presents its
own difficulties, even in the absence of disorder. Usually, for $d > 1$,
one has to resort to the limit of coordination number becoming large, with
interaction strength scaled suitably with coordination number to give a
nontrivial thermodynamic limit \cite{derrida}. The large-d limit in the
self-consistent field approximation for quantum-mechanical problems is
similar in spirit \cite{larged}.

The RFIM model on a Bethe lattice is special in that the zero-temperature
{\it nonequilibrium} response to a slowly varying magnetic field 
 can be determined exactly~\cite{DSS}. To be
precise, the average non-equilibrium magnetization in this model can be
determined exactly if the magnetic field is increased very slowly, from
$-\infty$ to $+\infty$, in the limit of zero temperature. It thus
provides a good theoretical model to study the slow relaxation to
equilibrium in glassy systems. The dynamics is governed by the existence
of many metastable states, with large energy barriers separating different
metastable states. We hope that this study of non-equilibrium response
in this model would help in  the more general problem of
understanding the statistical mechanics of metastable states in glassy
systems.

A brief summary of our results is as follows. We derive the exact self
consistent 
equations for the generating function of the avalanche size
distribution function $Q(x)$ 
on the Bethe lattice. This is a polynomial equation in $Q(x)$ and $x$,
in which the 
coefficients depend on the external field $h$, and the distribution of
the quenched 
random fields. We can solve these equations explicitly numerically and
thus determine the 
qualitative behavior of the distribution of avalanches for any
distribution of the 
quenched random fields. The behavior depends on the coordination
number $z$, and on the 
details of the distribution function.  We work out the distribution of
avalanches 
explicitly for a rectangular distribution of the quenched fields, for
the linear chain 
($z = 2$), and the 3-coordinated Bethe lattice. In both cases, one
finds only exponential 
decay.  We also studied other unimodal continuous distributions, {\it
e.g.} when the random field distribution is gaussian, or of the form 
$Prob(h_i)={1\over 2 \Delta} sech^2({h_i\over \Delta})$, also for
large $z$. We find that, for $z\ge4$,
 there is a regime of disorder strengths for which
the magnetization 
shows a jump-discontinuity ( ``first order transition''), but the
avalanche distribution, 
averaged over the hysteresis loop also shows a power law tail of the
form $s^{-5/2}$ 
(``critical fluctuations'').

The paper is organized as follows.  In section~\ref{model}, we define the
model precisely. In section~\ref{avalanche}, we briefly recapitulate the
derivation of self-consistent equations for the magnetization in our
model, and then use a similar argument to construct the generating
function for the avalanche distribution for arbitrary distribution of the
quenched random field. We set up a self-consistent equation for the
probability $Q_n$, that an avalanche propagating in subtree flips exactly
$n$ more spins in the subtree before stopping. The probability
distribution of avalanches is expressed in terms of this generating
function.  In section~\ref{RD}, we consider the special case of a
rectangular distribution of the random field. In this case, we explicitly
solve the self-consistent equations for Bethe lattices with coordination
numbers $z=2$ and $3$. However, this case is non-generic. For small
strength of disorder $\Delta$, the magnetization jumps from $-1$ to $+1$
at some value of the field, but for larger disorder, when the system shows
finite avalanches, there is no jump in magnetization  and the
distribution function 
decays exponentially for large $s$. In
section~\ref{GD}, we analyse the self-consistent equations to determine
the form of the avalanche distribution for some other unimodal
continuous distributions 
of the random field.  We find  that in each case for coordination
number $z\ge 4$, the 
magnetization  shows a first order jump discontinuity as a
function of the applied field at some field-strength $h_{disc}$, for weak
disorder. Just below $h = h_{disc}$, the avalanche distribution has a
universal $(-3/2)$ power-law tail.  Section~\ref{conclusion} contains a
discussion of our results, and some concluding remarks. Some algebraic
details of the analytical solution for the rectangular distribution of
quenched fields are relegated to two appendices.


\section{Definition of the  Model}
\label{model}

\begin{figure}
\begin{center}
\leavevmode
\psfig{figure=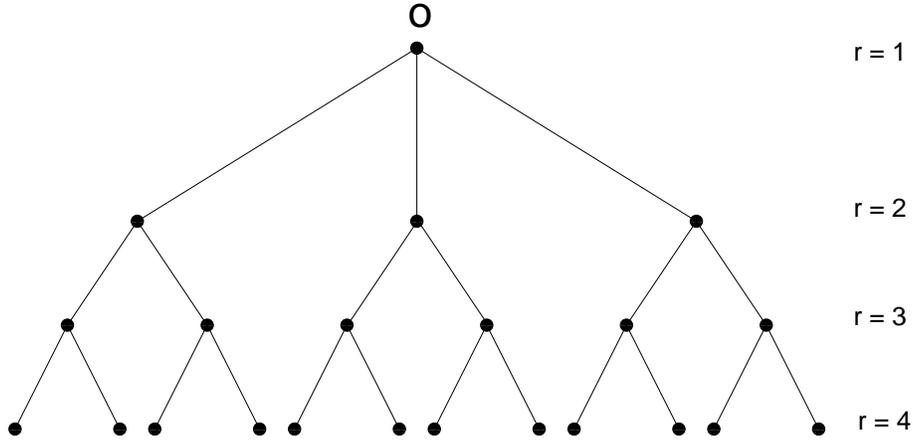,width=12cm}
\caption{A Cayley tree of coordination number 3 and  4 generations.}
\label{tree}
\end{center}
\end{figure}

We consider a uniform Cayley tree of $n$ generations where each
non-boundary site has a coordination number $z$ (see
Fig.~\ref{tree}). The first generation
consists of a single vertex. The $r$-th generation has $z(z-1)^{r-2}$ 
vertices for $r \geq 2$. 

The RFIM on this graph is defined as follows : At each vertex there is a
Ising spin $s_i=\pm 1$ which interacts with nearest neighbors through a
ferromagnetic interaction $J$. There are quenched random fields $h_{i}$ at
each site $i$ drawn independently from a continuous distribution $p(h_i)$.
The entire system is placed in an externally applied uniform field $h$.
The Hamiltonian of the system is

\be 
H=-J \sum_{<i,j>} s_is_j -\sum_{i}h_is_i -h\sum_{i}s_i \label{m.3} 
\ee

We consider the response of this system when the external field $h$ is
slowly increased from $-\infty$ to $+\infty$. 
We assume the dynamics to be  zero-temperature single-spin-flip
Glauber dynamics, {\it i.e.} a spin flip is allowed only if the process
lowers
energy. We assume that if the spin-flip is allowed, it occurs with a rate
$\Gamma$, which is much larger than the rate at which the magnetic field
$h$ is increased. Thus we  assume that all flippable spins relax
instantly, so that the spin $s_i$ is always  parallel to the net local
field $\ell_{i}$ at the site:

\be
s_i= sign(\ell_{i}) = sign( J \sum_{j=1}^{z} s_{j} + h_{i} + h)
\label{m.1}
\ee

We start with $h=-\infty$, when all spins are down and slowly increase
$h$. As we increase $h$, some sites where the quenched random field is
large positive will find the net local field positive, and will flip up.  
Flipping a spin makes the local field at neighboring sites increase, and
in turn may cause them to flip. Thus, the spins flip in clusters of
variable sizes. If increasing $h$ by a very small amount causes $s$ spins
to flip up together, we shall call this event an avalanche of size
$s$.  As the applied field increases, more and more spins flip up
until eventually all spins are up, and further increase in $h$ has no
effect. 


\section{The Self-Consistent Equations}
\label{avalanche}

The special property of the ferromagnetic RFIM that makes the analytical
treatment possible is this: Suppose we start with $h=-\infty$, and all
spins down at $t=0$. Now we change the field slowly with time, in such a
way that $h(t) \leq h(T)$, for all times $ t <T$. Then the 
configuration of spins at the final instant $t =T$ does not depend on the 
detailed time dependence of $h(t)$, and is the same for all histories, so
long as the condition $ h(t) \leq h(T)$ for all earlier times is obeyed.
In particular, if the maximum value $h(T)$ of the field was reached at
an earlier 
time $t_1$, then the configuration at time $T$ is exactly the same as
that at time 
$t_1$. This property is called the return point memory~\cite{sethna}.
We may choose to increase the field suddenly from $-\infty$
to $h(T)$ in a single step. Then, once the field becomes $h=h(T)$, several
spins would have positive local fields. Suppose there are two or more
such flippable sites. Then flipping any  one of them up can only 
increase the local field at other unstable sites, as all couplings are
ferromagnetic. Thus to reach a stable configuration, all such spins
have to be flipped, and {\it the final stable configuration reached is
the same, and independent of the order in which various spins are
relaxed}. This property will be called the abelian property of
relaxation. Using the symmetry between up and down spins, it is easy
to see that the abelian property also holds whether the new value of
field $h"$ is greater or less than its initial value $h'$ so long as
one considers transition from a stable configuration at $h'$ to a
stable configuration at $h"$. 
 
We first briefly recapitulate the argument of our earlier paper
\cite{DSS} which uses the abelian nature of spin-flips to
determine the mean magnetization for any field $h$ in the lower half of
the hysteresis loop by setting up a self-consistent equation.

Since the spins can be relaxed in any order, we relax them in this:  
first all the spins at generation $n$ ( the leaf nodes) are relaxed. Then
spins at generation $n-1$ are examined, and if any has a positive local
field, it is flipped. Then we examine the spins at generation $n-2$, and
so on. If any spin is flipped, its descendant are reexamined for possible
flips \cite{foot}. In this process, clearly the flippings of different
spins of the same generation $r$ are independent events.

Suppose we pick a site at random in the tree away from the
boundary, the probability that the local field at this site is
positive, given that exactly $m$ of its neighbors are up, is
precisely the probability that the local field $h_i$ at this site
exceeds  $[(z-2m)J-h]$. We denote this probability by $p_{m}(h)$. Clearly,

\be
p_{m}(h)=\int_{(z-2m)J-h}^{\infty} p(h_{i}) dh_{i}
\label{p_m}
\ee

Let $P^{(r)}(h)$ be the probability that a spin on the $n-r$-th generation
will be flipped when its parent spin at generation $n-r-1$ is kept down,
the external field is $h$, and each of its descendent spins has been
relaxed. As each of the $z-1$ direct descendents of a spin is
independently up with probability $P^{(r-1)}$, it is straightforward to
write down a recursion relation for $P^{(r)}$ in terms of $P^{(r-1)}$. For
$r >>1$, these probabilities tend to limiting value $P^\star$, which
satisfies the equation \cite{DSS}

\be 
P^\star (h)= \sum_{m=0}^{z-1} {z-1 \choose m} [P^\star (h)]^{m}
[1-P^\star (h)]^{z-1-m} ~p_{m}(h)  
\label{a.2} 
\ee

For the  spin at $O$, there are $z$ downward neighbors, and
the probability that it is up is given by

\be
Prob(s_O=+1|~h) = \sum_{m=0}^{z}  {z \choose m}  [P^\star (h)]^{m}
[1-P^\star (h)]^{z-m} ~p_{m}(h)  
\label{a.3}
\ee
  
Because all spins deep inside the tree are equivalent, $Prob(s_O=+1|~h)$
determines
the average magnetization for all sites deep inside the tree. Using
Eqs. (4-5), we can determine the magnetization for any value
of the external field $h$. This determines the lower half of the
hysteresis loop. The upper half is obtained similarly.

Now consider the state of the system at external field $h$, and all the
flippable sites have been flipped. We increase the field by a small
amount $dh$ till one more site becomes unstable. We would like to
calculate the probability that this would cause an `avalanche' of $n$ spin
flips. Since all sites deep inside are equivalent, we may assume the new
susceptible site is the site $O$.

It is easy to see that this avalanche propagation is somewhat like
propagation of infection in the contact process on the Bethe lattice. The
`infection' travels $\it downwards$ from the site $O$ which acts as the
initiator of infection. If any site is infected, then it can cause
infection of some of its descendents. If the descendent spin is already
up, it cannot be flipped; such sites act as immune sites for the infection
process. If the descendent spin is down, it can catch infection with a
finite probability. Furthermore, this probability does not depend on
whether the other  
`sibling' sites catch infection. 
Infection of two or more descendents of
an infected site are uncorrelated events. Thus, we can expect to find the
distribution of avalanches on the Bethe lattice, as for the size
distribution of percolation clusters on a Bethe lattice
\cite{percolation}. However, a precise description in terms of the
contact process is 
complicated, as here the infection spreads in a correlated background
of `immune' 
(already up ) spins, and the probability that a site catches infection
does depend on the 
number of its neighbors that are already up.

\begin{figure}
\begin{center}
\leavevmode
\psfig{figure=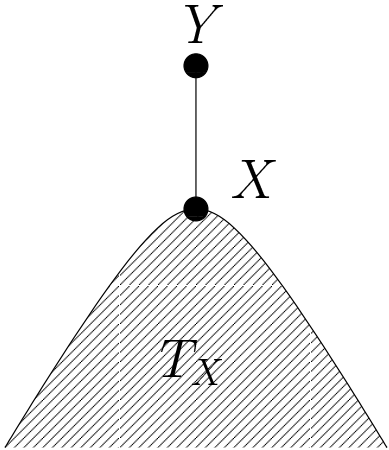}
\caption{A sub-tree $T_X$ formed by $X$ and its descendents. The
sub-tree is rooted at $X$ and $Y$ is the parent spin of $X$.}
\label{sub-tree}
\end{center}
\end{figure}

We start with the initial configuration of all spins down. Now increase
the external field to the value $h$. Consider a site $X$ at some generation $r
>1$ of the Cayley tree [Fig.~\ref{sub-tree}]. We call the subtree
formed by $X$ and its 
descendents $T_X$, the subtree rooted at $X$. We keep its parent spin $Y$
at generation $r-1$ down, and relax all the sites in $T_X$ at the
uniform field 
$h$. If $X$ is far away from the boundary, the probability that spin
at $X$ is up is $P^\star (h)$. 
The conditional probability that spin at a descendant of $X$ is up,
given that the spin at $X$ is down is also $P^\star (h)$.  We measure the
response
of $T_X$ to external perturbation by forcibly flipping the spin at $Y$ (
whatever the local field there) and see how many spins in this subtree
flip in response to this perturbation. Let $Q_n$ be the probability that
the spin at $X$ was down when $Y$ was down {\it and} $n$ spins on the
subtree $T_X$ flip up if $S_Y$ is flipped up. Here allowed values of $n$
are $0,1,2,\ldots$. Clearly, we have

\be
P^\star + \sum_{n=0}^{\infty} Q_n =1
\ee

\noi We define 
\be
Q(x)= \sum_{n=0}^{\infty} Q_n x^n
\label{def.Q}
\ee

Clearly, $Q(x=0)=Q_0$ and $Q(x=1) = 1-P^\star$. It is straight forward to
write the self-consistent equation for $Q(x)$.  Let us first relax all
spins on $T_X$ keeping $X$ and $Y$ down. The probability  that exactly
$m$ the descendents of $X$ are turned up in this process be denoted by
$Pr(m)$. Clearly

\be
Pr(m) =  {{z-1} \choose m} {P^\star}^m (1 -P^\star)^{z-1-m}    
\ee

For a given $m$, the conditional probability that local field at $X$ is
such that spin remains down, even if $Y$ is turned up is $1-p_{m+1}$.
Summing over $m$, and using the expression for $Pr(m)$ above, we get 

\be
Q_{0} = \sum_{m=0}^{z-1}{{z-1} \choose m} {P^\star}^m (1
-P^\star)^{z-1-m} [1- p_{m+1}] 
\label{Q_0}
\ee

We can write down an expression for $Q_{1}$ similarly. In this case, if
$m$ of the direct descendents of $X$ are up when $Y$ is down, the local
field at all the remaining $z-1-m$ direct descendents must be
such that they remain down even if $X$ is flipped up.
This probability is $ { {z -1 } \choose m } {P^*}^m Q_{0}^{z-1-m}$. The
local quenched field at $X$ must satisfy $ (z-2m)J -h > h_X > (z - 2m -2)J
-h$. The probability for this to occur is $p_{m+1}-p_{m}$.
Hence we get

\be
Q_{1}=\sum_{m=0}^{z-1} (p_{m+1} - p_{m}){{z -1 } \choose m}  {P^\star}^m
~Q_{0}^{z-1-m}
\ee

The equation determines $Q_{n}$ for higher $n$ can be written down
similarly. It 
only involves the probabilities $Q_{m}$ with $m < n$ for the descendent
spins. These recursion equations are expressed more simply in terms of the
generating function $Q(x)$. It is easily checked that the self-consistent
equation for $Q(x)$ is

\be
Q(x)= Q(x=0) + x  \sum_{m=0}^{z-1} {{z-1} \choose m}(p_{m+1}-p_m)
{P^\star}^m ~Q(x)^{z-1-m} 
\label{Q(x)}
\ee

This is a polynomial equation in $Q(x)$ of degree $z-1$, whose
coefficients are functions of $h$ through $P^\star(h)$ and $p_m(h)$. It is
easily checked that for
$x=1$, the ansatz  $Q(x=1)=1- P^\star$ satisfies the equation, as it
should.  
To determine $Q(x)$ for any given external field $h$, we have to first
solve the self-consistent equation for $P^\star$ [ Eq.~\ref{a.2}]. This then
determines $Q(x=0)$ using Eq.~\ref{Q_0}, and then, given $P^\star$ and
$Q(0)$, we 
solve for $Q(x)$ by solving the $(z-1)$-th degree polynomial equation
Eq.~\ref{Q(x)}.  
 
Finally, we express the relative frequency of avalanches of various sizes
when the external field is increased from $h$ to $h + dh$ in terms of
$Q(x)$. Let $G_s(h)dh$ be  the probability that
avalanche of size $s$
is initiated at $O$.  We also define the generating
function $G(x|h)$ as

\be
G(x|h)= \sum_{s=1}^{\infty} G_s(h) x^s
\label{def.G}
\ee

Consider first the calculation of $G_s(h)$ for $s=1$. Let the number of
descendents of $O$ that are up at field $h$ be $m$. For the spin at site
$O$ to be down at $h$ , but flip up at $ h+dh$, the local field $h_{O}$
must satisfy $ [(z-2m)J-(h +dh)] < h_{O} < [(z-2m)J-h]$. This occurs with
probability $p(zJ-2mJ-h)dh$. Each of the $(z-m)$ down neighbors of $O$
must not flip up, even when $s_O$ flips up. The conditional probability of
this event is $Q_0^{z-m}$.  Multiplying by the probability that 
$m$ neighbors are up, we finally get

\be 
G_1(h) = \sum_{m=0}^{z} { z \choose m} {P^\star}^m ~{Q_0}^{z-m} ~p(zJ-2mJ-h)
\ee  

Arguing similarly, we can write the equation for $G_s(h)$ for $s=2,
3 $ etc. These equations simplify considerably when expressed in terms of the generating
function $G(x|h)$, and we get

\be
G(x|h) = x\sum_{m=0}^{z} { z \choose m} ~{P^\star}^m ~{Q(x)}^{z-m}
~p(zJ-2mJ-h)
\label{G(x|h)}
\ee

In numerical simulations, and experiments, it is much easier to measure
the avalanche distribution integrated over the full hysteresis loop.
To get the probability that an avalanche of size $s$ will be initiated
at any given site $O$ in the interval when the external field is increased
from $h_1$ to $h_2$, we just have to integrate $G(x|h)$ in this range.
For any $h$, the value of $dG/dx$ at $x=1$ is proportional to the mean size
of an avalanche, and thus to the average slope of the hysteresis loop 
at that $h$. 

\section{Explicit calculation for the Rectangular distribution}
\label{RD}

While the general formalism described in the previous section  can be used
for any distribution, and any coordination number, to calculate the
avalanche distributions explicitly, we have to choose some specific form
for the probability distribution function. In this section, we shall 
consider the specific choice of a rectangular distribution : The
quenched random field is uniformly distributed between $-\Delta$ and
$\Delta$, so that

\be
p(h_i) = \frac{1}{2 \Delta}~,   \mbox{~~for }~~ -\Delta \leq h_i \leq
\Delta 
\ee

In this case, the cumulative probabilities $p_m(h)$ become piecewise
linear functions of $h$, and $h$-dependence of the distribution is easier
to work out explicitly. We shall work out the distributions  for
the linear chain ( $z = 2$), and the 3-coordinated Bethe lattice. 

\subsection{The Linear Chain}  
\label{z.2}
\begin{figure}[htbt]
\begin{center}
\leavevmode
\psfig{figure=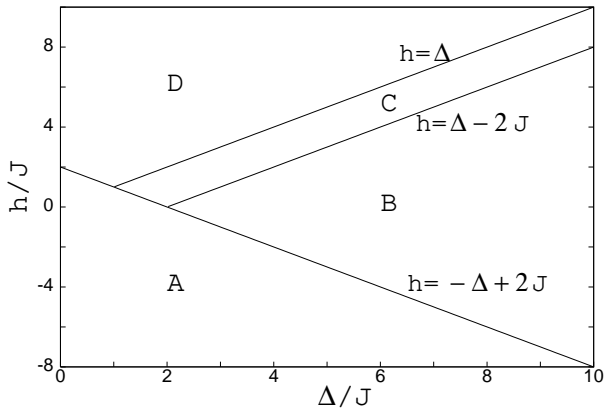,height=6cm,angle=-90}
\caption{ Behavior of RFIM  in the magnetic field - disorder
($h-\Delta$) plane for a linear chain. The regions A-D correspond to
qualitatively  different responses. In region A all spins are down and
in region D all are up. The avalanches of finite size occur in region
B and C.} 
\label{phase2}
\end{center}
\end{figure}

The simplest illustration  is for a linear chain. In this case the
self-consistent equation, for the probability
$P^\star$ [ Eq.~\ref{a.2} ] becomes a linear equation. This is easily
solved, and explicit expressions for $Q_0$, and $Q(x)$ are obtained
(see Appendix~\ref{appendixA}). The different regimes showing
different qualitative behavior of the hysteresis loops are shown in
Fig.~\ref{phase2} 

For $h < 2J -\Delta$ (region A), all the spin remain down.  For $h >
\Delta$, all spins are up (region D). For $\Delta <J$, we get a
rectangular loop and the magnetization jumps discontinuously from $-1$ to
$+1$ in a single infinite avalanche, and we directly go from region A to D
as the field is increased. For $\Delta >J$, we get nontrivial hysteresis
loops.

\begin{figure}
\begin{center}
\leavevmode
\psfig{figure=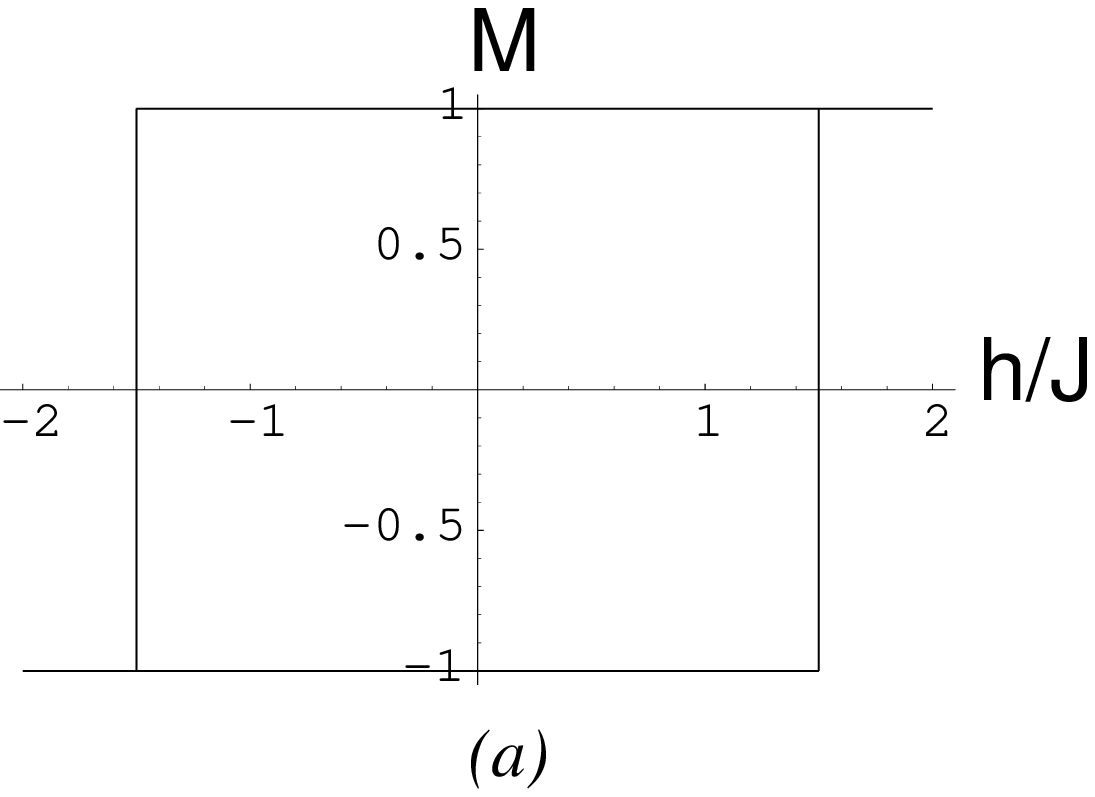,width=6cm}
\psfig{figure=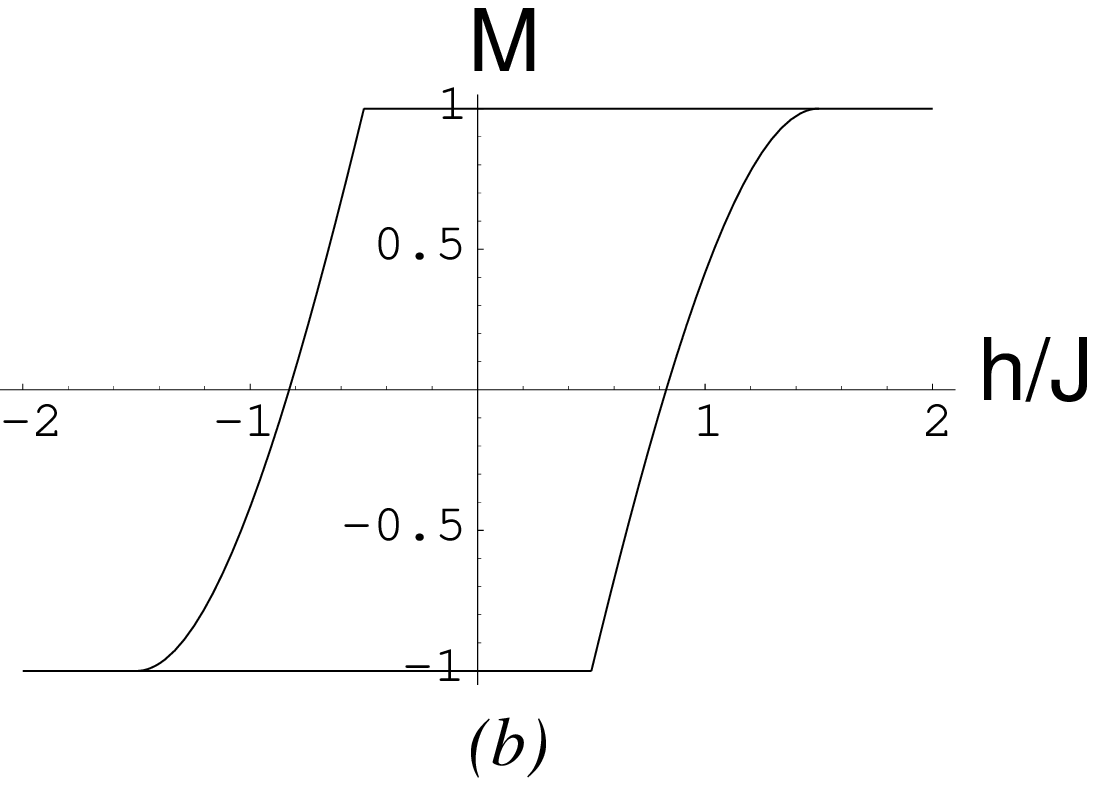,width=6cm}
\leavevmode
\psfig{figure=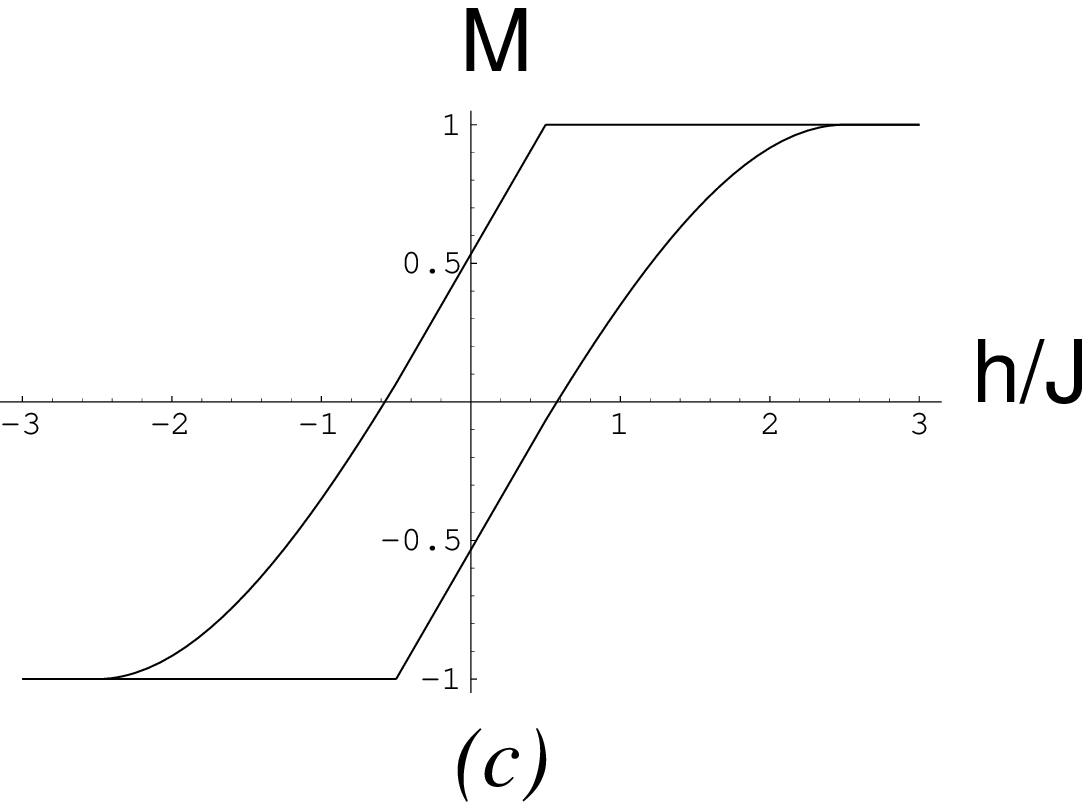,width=6cm}
\caption{Hysteresis loops for the linear chain for the
rectangular distribution of quenched fields with different widths 
{\it (a)} $\Delta/J=0.5$, {\it (b)} $\Delta/J=1.5$ and {\it (c)}
$\Delta/J=2.5$} 
\label{m2}
\end{center}
\end{figure}

The hysteresis loops for different
values of $\Delta =0.5, 1.5$ and $2.5$ are shown in Fig.~\ref{m2}.
If $\Delta$ is sufficiently large ($ \Delta > J$), we find that the mean
magnetization is a precisely linear function of the external field for a
range of values of the external field $h$ (region B in Fig. 2). For larger 
$h$ values, the magnetization shows saturation effects, and is no longer
linear ( region C). 

The explicit forms of the generating function $Q(x)$ are given in the
Appendix~\ref{appendixA}. We find that in region B, the function
$Q(x)$ is independent of 
the applied field $h$. The distribution function $G_s(h)$ has a simple
dependence on $s$ of the form

\be
G_s(h)= A_1 s  \left(\frac{J}{\Delta}\right)^{s},
\label{Gs.B.z2}
\ee
where $A_1$ is a constant, that depends only on $J/\Delta$, and does not depend on
$s$ or $h$
\be
A_1= {1 \over 2 \Delta}~\frac{(1-J/\Delta)^2}{(J/\Delta)}
\ee

In region C, the mean magnetization is a nonlinear function of $h$. But
$Q(x)$ is still a rational function of $x$. From the explicit functional form 
of $Q(x)$ and $G(x|h)$ are given in the appendix~\ref{appendixA}, we find that
$G_s(h)$ is of the form 
\be
G_s(h) =[ A_1' s + A_2'] \left({J\over \Delta}\right)^{s}
                    , ~~{\mbox for}~~ s \geq 2.
\ee
Here $A_1'$ and $A_2'$   have no
dependence on $s$ but are explicit functions 
of $h$.

Integrating over $h$ from $-\infty$ to $\infty$ we get the integrated
avalanche distribution  $D_s$,
\be
D_s = \int_{-\infty}^{\infty} G_s(h) dh
\ee
It is easy to see from above that the integrated distribution $D_s$ also
has the form

\be
D_s = [ A_2 s + B_2] \left({J\over \Delta}\right)^s, {\rm for}~~ s \ge 2
\ee
where the explicit forms of the coefficients $A_2$ and $B_2$ are given in
the Appendix~\ref{appendixA}.

\subsection{The Case $z = 3$}
\label{z3}
\begin{figure}[htbt]
\begin{center}
\leavevmode
\psfig{figure=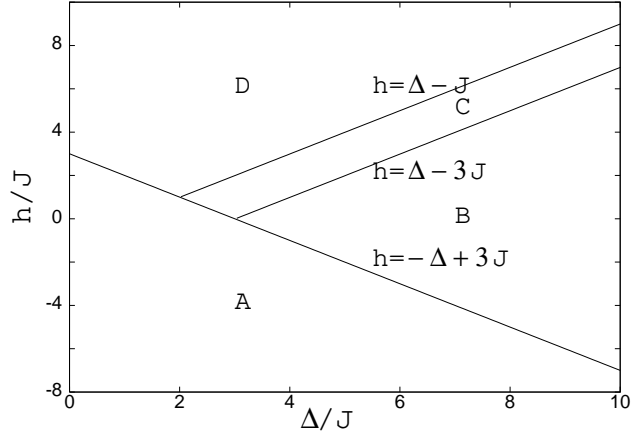,height=7cm,angle=-90}
\caption{ Behavior of RFIM  in the magnetic field - disorder
($h-\Delta$) plane for Bethe lattice of coordination number 3. The
qualitative behavior in different regions A-D is similar to that of
a linear chain (Fig.~\ref{phase2}).}
\label{phase3}
\end{center}
\end{figure}

The analysis for the case $ z=3$ is very similar to the linear case. In
this case,
the self-consistent equation. for $P^\star(h)$ [ Eq.~\ref{a.2} ]
becomes a quadratic equation. The qualitative behavior of solution is
very similar to the earlier case. Some details are given in
Appendix~\ref{appendixB}. 
We again get regions A-D as before, but  the boundaries 
are shifted a bit, and are shown in  Fig.~\ref{phase3}.
As before, in region B, the average
magnetization
is a linear function of $h$, and  the avalanche distribution is
independent of $h$.  

We find that in  regime B, the distribution of
avalanche sizes is given by

\be
G_s(h) =N
\left[\frac{(2s)!}{(s-1)!(s+2)!}\right]
	(1-J/\Delta)^{s}\left({J\over\Delta}\right)^{s}  
\label{Gs.B.z3}
\ee
where $N$ is a normalization constant given by
\be
N={3 \over 2 \Delta} (1 - J/\Delta)^2 { 1 \over (J/\Delta)}
\ee

\noi It is easy to see that  for large $s$, $G_s(h)$ varies as 

\be
G_s \sim s^{-\frac{3}{2}} \kappa^s
\ee
where
\be
\kappa = 4 ( 1 - J/ \Delta) (J/\Delta)
\ee
In region B,   $J/\Delta$ is always less than $1/3$, and so this
function always has an exponential decay for large $s$.

In the region C,  we find that the avalanche distribution is of the form

\be
G_s(h)= N'
\left[\frac{(2s)!}{(s-1)!(s+2)!}\right] \kappa^s  
\label{Gs.C.z3}
\ee
where $N'$ is a normalization constant independent of $s$, and $\kappa$ is a
a cubic polynomial in the external field $h$:
 
\bea
\kappa = {1\over 8(1-2J/\Delta)^2}
	&&\left[ 
	\left\{9-53(J/\Delta)+119(J/\Delta)^2-107(J/\Delta)^3\right\}
	 \right. \nn \\
	&& \left.  +\left\{-5+10(J/\Delta)+11(J/\Delta)^2\right\}(h/\Delta)
	+\left\{3-9(J/\Delta)^2\right\}(h/\Delta)^2 +(h/\Delta)^3
	\right]
\eea
As $\kappa$ is not a very simple  function of
$h$, explicit expressions for the integrated distribution $D_s$ are
hard to write down.  


\section{General Distributions } 
\label{GD}

The analysis of the previous section can, in principle, be extended to
higher coordination numbers, and other distributions of random fields.
However, the self-consistent equations become cubic, or higher order
polynomials. In principle, an explicit solution is possible 
for $z \le 5$, but it is not
very instructive. However, the qualitative behavior of
solutions is easy to determine, and is the same for all $ z \ge 4$. 
We shall take $z=4$ in the following for simplicity.  Since we only study
the
general features of the self-consistent equations, we need not pick a
specific form for the continuous distributions of random field
distribution $p(h_i)$. We shall only assume that it has a single
maximum around zero and asymptotically go to zero at $\pm \infty$. 

For small width ( $\Delta$ )
of the random field distribution {\it i.e.} for weak disorder the
magnetization shows a jump discontinuity as a function of the external
uniform field , which disappears for a larger values of $\Delta$
~\cite{DSS}. For fields $h$ just lower than the value where the jump
discontinuity occurs, the slope of the hysteresis curves is large, and
tends to infinity as the field tends to the value at which the jump
occurs. This indicates that  large avalanches  are more likely just
before the first
order jump in magnetization.

\begin{figure}[htbt]
\begin{center}
\leavevmode
\psfig{figure=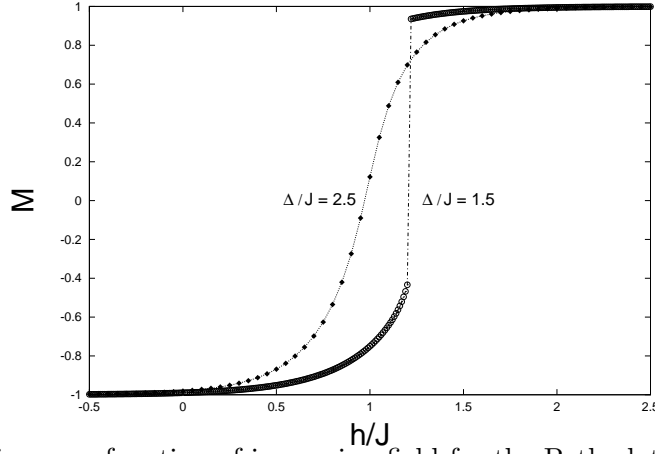,height=6cm,angle=-90}
\caption{Magnetization as a function of increasing field for the Bethe
lattice
with $z=4$ and the random field distribution given by Eq.~\ref{sech}.}
\label{mag}
\end{center}
\end{figure}

For $z=4$, the self-consistent equation for  $P^\star(h)$ 
[ Eq.~\ref{a.2} ] is  cubic 
\be
a P^{\star 3}+b P^{\star 2} +c P^\star+d=0
\label{z4.1}
\ee
where $a, b, c$ and $d$ are functions of the external field $h$, 
expressible in terms of the cumulative
probabilities $p_i,i=0$ to $3$,
\bea
&& a=p_3-3p_2+3p_1-p_0 \nonumber \\
&&b=3p_2-6p_1+3p_0 \nonumber \\
&&c=3p_1-3p_0-1 \nonumber \\
&&d=p_0 \nonumber 
\eea

\begin{figure}[htbt]
\begin{center}
\leavevmode
\psfig{figure=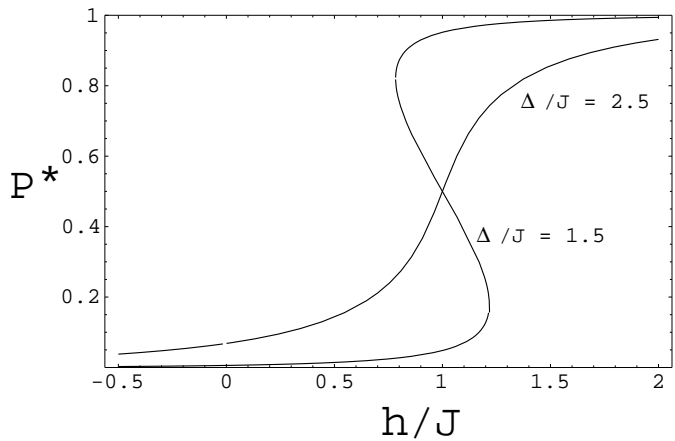,height=6cm}
\caption{Variation of $P^\star(h)$ with $h$  for the Bethe lattice with
$z=4$, and the  random field distribution given by Eq.~\ref{sech}.}
\label{root}
\end{center}
\end{figure}

This equation will have $1$ or $3$ real roots, which will vary with $h$.
We have shown this variation for the real roots which lie between 0 and 1
in Fig.~\ref{root} for the case where $p(h_i)$ is a simple
distribution

\be
p(h_i) = \frac{1}{2 \Delta} sech^2( h_i/\Delta)
\label{sech}
\ee
We have also solved numerically the self-consistent equation for
$P^\star$ for 
other choices of $p(h_i)$, like the gaussian distribution, and for
higher $z ( = 4,5,6 )$. In each case we find that the qualitative
behavior  of the solution is very similar.
Note that the rectangular distribution discussed in the previous section
is very atypical in that both the coefficients $a$ and $b$ vanish for an
entire range of values of $h$.

In the generic case, we find two qualitatively different behaviors: For
larger values of $\Delta$, there is only one real root for any $h$ . For
$\Delta$ sufficiently small, we find a range of $h$ where there are $3$
real solutions.  There is a critical value $\Delta_c$ of the width which
separates these two behaviors. For the particular distribution chosen
$\Delta_c \simeq 2.10382$.

In the first case, the real root is a continuous function of $h$, and
correspondingly, the magnetization is a continuous function of $h$. This is
the case corresponding to $\Delta =2.5$ in Fig.~\ref{root}.

For smaller  $\Delta < \Delta_c$, for
large $\pm h$  there is only one root , but in the intermediate region
there are three roots. The typical variation is shown for $\Delta =
1.5$ in
Fig.~\ref{root}. In the increasing 
field the probability $P^\star(h)$ initially takes the smallest root.
As $h$ increases , at a  value $h=h_{disc}$ , the
middle and the lower roots become equal and after that both disappear
from the real plane . At $h=h_{disc}$ the probability $P^\star(h)$ jumps to the
upper root .  Thus for $\Delta < \Delta_c$ there is a discontinuity in
$P^\star (h)$ which gives rise to a first order jump in the magnetization
curve . 

The field $h_{disc}$ where the discontinuity of magnetization occurs,
is determined by the condition that for this value of $h$, 
the cubic equation [ Eq.~\ref{z4.1} ] has two equal roots. The value
of $P^\star$ at this point, denoted by $P^\star_{disc}$, satisfies the
equation 
\be 
3 a_0 P^{\star 2}_{disc}+2 b_0 P^\star _{disc}+c_0=0 
\label{z4.2}
\ee
where $a_0, b_0$ and $c_0$  are the values of $a, b$ and $c$ at
$h=h_{disc}$.  

We now determine the behavior of the avalanche generating function
$G_s(h)$ for large $s$ and $h$ near $h_{disc}$. The behavior for large
$s$ corresponds to $x$ near $1$. So we write $x = 1- \delta$, with
$\delta$ small, and $h = h_{disc} -\epsilon $. Near $h_{disc}$, $a, b, \ldots$
vary linearly with $\epsilon$ and 
\be
P^\star \approx P^\star _{disc} - \alpha \sqrt{\epsilon} + O(\epsilon)
\ee
where $\alpha$ is a numerical constant.

Since $Q(x=1)=1-P^\star(h)$ , if $x$ differs slightly from unity $Q(x)$ also
differs
from $ 1 - P^\star(h)$ by  a  small amount. Substituting 
$x=1-\delta$ and $Q(x=1-\delta)=1-P^\star-F(\epsilon, \delta)$ in
the self-consistent equation for $Q(x)$ [Eq.~\ref{Q(x)}], 
where both  $\delta$ and  $F$ are small, using 
Eq.~\ref{z4.2}, we get 
to lowest order in $\delta$, $\epsilon$ and $F$
\be
F^2 + \beta \sqrt{\epsilon} F - \gamma^2 \delta = 0
\label{z4.5}
\ee
where $\beta$ and $\gamma$ are some constants.
Thus, to lowest orders in $\epsilon$ and $\delta$, $F$ is given by
\be
F = (1/2)[ \sqrt{\beta^2 \epsilon + 4 \gamma^2 \delta} - \beta \sqrt{\epsilon}]
\ee

Thus $Q(x)$ has leading square root singularity at
$x=1+\frac{\beta^2 \epsilon}{4 \gamma^2}$.
Consequently, $G(x|h)$ will also show a square root
singularity $x=1+\frac{\beta^2 \epsilon}{4 \gamma^2}$.
This implies that  the Taylor expansion coefficients
$G_s(h)$ vary as

\be
G_s(h) \sim  s^{-\frac{3}{2}} \left(1+\frac{\beta^2 \epsilon}{4
\gamma^2}\right)^{-s},  
~~~~\mbox{for large $s$.}
\label{z4.7}
\ee

At $\epsilon=0$, we get
\be
G_s(h_{disc}) \sim  s^{-\frac{3}{2}} 
\ee
Thus at $h=h_{disc}$ the avalanche distribution has a power law tail.

To calculate the integrated distribution $D_s$, we have to integrate
Eq.~\ref{z4.7} over a range of $\epsilon$ values. For large $s$, only
$\epsilon < {\gamma^2 \over \beta^2 s}$ contributes significantly to
the integral, and thus we get 
\be
D_s \sim s^{-{5\over 2}} \mbox{~,~~~~for large} ~~s.
\ee
Thus the integrated distribution shows a robust $(-5/2)$ power law for
a range of disorder strength $\Delta$. 


\section{Discussion}
\label{conclusion}

In this paper, we set up exact self-consistent equations for the avalanche
distribution function for the RFIM on a Bethe lattice. We were able to
solve these equations explicitly for the rectangular distribution of the
quenched field, for the linear chain $z=2$, and the 3-coordinated Bethe
lattice. For more general coordination numbers, and general
continuous distributions of random fields, we argued that for very large
disorder, the avalanche distribution is exponentially damped, but for
small disorder, generically, one gets  a jump in magnetization,
accompanied by a square-root singularity. For field-strengths just below
corresponding to the
jump discontinuity,  the avalanche distribution function has a
power-law tail of the form $s^{-3/2}$. The integrated avalanche
distribution then varies as  $s^{-5/2}$  for large $s$.

Some unexpected features of the solution deserve mention. Firstly,
we find that the behavior of the self-consistent equations for $z=3$
is qualitatively different from that for $z>3$. The behavior for the
linear chain ($z = 2$) is, of course, expected to be different from higher
$z$. One usually finds same behavior for all $z >2$. Mathematically, the
reason for this unusual dependence is that the mechanism of two real
solutions of the polynomial equation merging, and both becoming unphysical
(complex) is not available for $z=3$. Here the self-consistency equation
is  a quadratic, and from physical arguments, at least one of the roots
must be real. That  a Bethe lattice may show non-generic behavior for low
coordination numbers has been noted earlier by  Ananikyan {\it et al} in their
study of the 
Blume-Emery-Griffiths model on  a Bethe lattice. These authors observed
that  the qualitative behavior for $z < 6$ is different from that for  $z
\geq 6$ \cite{ananikyan}.

The second point we want to emphasize is that here we find that the
power-law tail in the distribution function is accompanied by the
first-order jump in magnetization. Usually, one thinks of critical
behavior and first-order transitions as mutually exclusive, as first-order
jump pre-empts a build-up of long-ranged correlations, and all
correlations remain finite-ranged across a first -order transition. This
is clearly not the case here. In fact, the power-law tail in the avalanche
distribution disappears, when the jump disappears. A similar situation
occurs in equilibrium statistical mechanics in the case of a Heisenberg
ferromagnet below the critical temperature. As the external field $h$ is
varied across zero, the magnetization shows a jump discontinuity, but in
addition has a cusp singularity for small fields~\cite{parisi}. But in this
case the power-law tail is seen on {\it both sides of the transition}.

Note that for most values of disorder, and the external field, the
avalanche distribution is exponentially damped. We get robust power law
tails in the distribution, only if we integrate the distribution over the
hysteresis cycle across the magnetization jump. But, in this case, the
control parameter $h$  is swept across a range of values, in
particular across a (non-equilibrium) phase transition point! In this
sense, while no 
explicit fine-tuning is involved in an experimental setup, this is not a
self-organized critical system in the usual sense of the word.
Recently P\'{a}zm\'{a}ndi {\it et al} have argued that the hysteretic
response of the 
Sherrington-Kirkpatrick model to external fields at zero temperature
shows 
self-organized criticality for all values of the field \cite{pazmandi}.
However, this seems to be because of the presence of infinite-ranged 
interactions in that model.

The treatment of this paper may be extended to the site-dilution case
discussed by Tadi\'{c}~\cite{tadic}. From the structural stability of
the mechanism which 
leads to the cusp singularity just before the jump-discontinuity in
magnetization, it is clear that in our model, introduction of
site dilution would not change the qualitative behavior of solutions.

A general question concerns the behavior of the avalanches for more
general probability distributions. Clearly, if $p(h_i)$ has a discrete
part, it would give rise to 
jumps in $p_i$ as a function of $h$, and hence give rise to several
jumps in the hysteresis loop. These could preempt the cusp singularity
mechanism which is responsible for the power-law tails.
If the distribution $p(h_i)$ is continuous, but multimodal, then it is
possible to have more than one first order jump in the
magnetization \cite{z6}. This is confirmed by explicit calculation in some
simple cases. If $p(h_i)$ has
power-law singularities, these would also lead to power-law singularities
in $p_i$, and hence in $P^\star(h)$.  Even for purely continuous
distributions, the merging of two roots as the magnetic field varies
need not always occur. For example, it is easy to check that for the
rectangular distribution, even for $z \ge 4$, we do not get a power law
tail for any value of $\Delta$. The precise conditions necessary for
the occurrence 
of the power-law tail are not yet clear to us.

Finally, we would like to mention some open questions. Our analysis
relied heavily on the fact that initial state was all spins down. Of
course, we can start with other initial conditions. It would be
interesting to set up self-consistent field equations for them. In
particular, the behavior the return loop, when the external field is 
increased from $-\infty$ to some value $h_1$, and then decreased to a
lower value $h_2$ seems an interesting quantity to determine. Another 
extension would be to make the rate of field-sweep comparable to the
single-spin flip rate (still assuming T=0 dynamics). This would mean some
large avalanches in different parts of the sample could be evolving
simultaneously. Then one could
study the sweep-rate
dependence of the hysteresis loops, and  the frequency dependence of
the Barkhausen noise spectra. This is
perhaps of some relevance in real experimental data, and would also make
contact with other treatments of Barkhausen noise that focus on the domain
wall motion.
 
We thank M. Barma and N. Trivedi for their useful comments on the
manuscript.
DD would like to thank the Physics Department of North Eastern Hill
University, for hospitality during  a visit there.

\appendix
\section{}  
\label{appendixA}
\noi For the case of a  linear chain, the
self-consistent equation, for the probability
$P^\star$ [ Eq.~\ref{a.2} ] is a linear equation,  whose solution is  ,
\be
P^\star (h) = \frac{p_0}{[ 1 - ( p_1 - p_0 ) ] }
\label{z2.P*}
\ee

\noi For $h < 2J -\Delta$, $p_0$ is zero, and hence $P^\star (h)$ is
zero, and all the spin remain down (region A in Fig.~\ref{phase2}).

\noi For $h > 2J-\Delta$, and $\Delta < J$, $p_1$ is $1$ whenever $p_0$ is
nonzero. Then from Eq.~\ref{z2.P*}, $P^\star (h)$ becomes $1$. Thus, for
$\Delta <J$, we get a rectangular loop and the system changes from all
spins down to all spins up state in a single big avalanche.

\noi For $ \Delta > J$, $p_1 -p_0$ equals $J/\Delta$ and is independent of $h$,
in the range $ 2J- \Delta < h < \Delta$. Thus $P^\star (h)$ is  a linear
function of $h$ in this range, increasing from $0$ to $1$.

Defining 
\be
\epsilon = \frac{1}{2} ( 1 + \frac{h}{\Delta} - \frac{2 J}{\Delta} )
\ee
we obtain the expression for $P^\star$ as 
\be
P^\star(h) =\cases{  0 & {for $\epsilon <  0$ } \cr
       \frac{\epsilon}{1 - J/\Delta} 
	&  {for $0 \le \epsilon \le 1 - J/\Delta$ } \cr
        1 & {for  $\epsilon > 1- J/\Delta$ } 
}
\ee
Using Eq. (\ref{Q_0}), the expression for $Q_0$ is,
\be
Q_0 = (1-p_1) - (p_2-p_1)P^\star (h)
\ee
The generating function $Q(x)$ obtained from the self-consistent
equation [Eq.\ref{Q(x)}] is,
\be
Q(x)=\frac{Q_0 +x P^\star (p_2 -p_1)}{1-x (p_1 -p_0)}
\ee
and the generating function $G(x|h)$ given by Eq.~\ref{G(x|h)} becomes ,
\be
G(x|h)=x\left\{ [Q(x)]^2 p(2J-h) +2 P^\star [Q(x)] p(-h) +P^{\star 2}
p(-2J-h)\right\}
\ee
Now  if $\Delta > 2J$, and $ -\Delta + 2J < h < \Delta - 2J$ ( region B in
Fig.~\ref{phase2} ),
\bea 
&&(p_2-p_1)=(p_1-p_0)=J/\Delta , \nonumber \\
&& p(2J-mJ-h)=\frac{1}{2\Delta} \mbox{ ~~~for ~~$m=0,1,2 $} \nonumber   \\ 
\mbox{and~~~~ } && P^\star+Q_o=1-J/\Delta \nonumber
\eea
Thus  
\be
Q(x)=\frac{1-(J/\Delta)}{1-(J/\Delta) x}-P^\star 
\ee
and
\be
G(x|h)= \frac{x}{2\Delta}\left[P^\star+Q(x)\right]^2 
	=\frac{x}{2\Delta} ~{ (1-J/\Delta)^2 \over (1-xJ/\Delta)^2}
\ee
Expanding $G(x|h)$ in powers of $x$, we get the probability distribution of
avalanches in region B  given by Eq.~\ref{Gs.B.z2} of sec.~\ref{z.2}.

In the region C, $p_2$ saturates to value $1$,
$p(-2J-h)$ becomes zero and $(p_2 - p_1)$ becomes
$(1-J/\Delta-\epsilon)$ . Thus we get ,

\be
Q_0 = \frac{(1 -J/\Delta -\epsilon)^2}{(1 - J/\Delta)}
\ee
In terms of $P^\star$ and $Q_0$ we get 
\be
Q(x)=\frac{Q_0 + x P^\star [1-2 (J/\Delta) -\epsilon]}{1-(J/\Delta) x}
\label{q}
\ee
and
\be
G(x|h)= \frac{x}{2\Delta}\left\{ [P^\star + Q(x)]^2 - P^{\star
 	    2}\right\}
\label{p}
\ee
Expanding $G(x|h)$ in powers of $x$
we get , in region C
\be
G_1(h) = \frac{1}{2\Delta}\left[(P^\star + Q_0)^2 - P^{\star 2}\right]
\ee
and
\be
G_s(h) =[ A_1' s + A_2'] \left({J\over \Delta}\right)^{s}
                    , ~~{\mbox for}~~ s \geq 2.
\ee
Here $A_2$ and $B_2$  have no
dependence on $s$ but are explicit functions 
of $h$
\bea
A_1' & =& {1\over 2 \Delta} \left[{1\over (J/\Delta)} (P^\star + Q_0)^2
+{1\over (J/\Delta)^2}  (P^\star + Q_0) P^\star (1 - {2J\over \Delta}
-h/\Delta) \right. \nn \\
&& ~~~~~~~\left. + {1\over 4 (J/\Delta)^3}{P^\star}^2 (1 -{2J\over \Delta}
-h/\Delta)^2\right]   \nonumber \\
A_2 '& =& {-1\over \Delta)}\left[
 {1\over 2(J/\Delta)^2}(P^\star + Q_0) P^\star (1 - {2J\over \Delta}-h/\Delta)
+{1\over 4(J/\Delta)^3}{P^\star}^2 (1 -{2J\over \Delta} -h/\Delta)^2\right]
 \nonumber
\eea
Integrating over $h$ from $-\infty$ to $\infty$ we get the integrated
avalanche distribution  $D_s$,
\be
D_s = \int_{-\infty}^{\infty} G_s(h) dh
\ee
where
\be
D_{1}= \frac{1}{(1-J/\Delta)^{2}} \left[
1-6\left({J\over \Delta}\right) +14 \left({J\over \Delta}\right)^{2} -
\frac{46}{3} \left({J\over \Delta}\right)^{3} 
+\frac{47}{6} \left({J\over
\Delta}\right)^{4}-\frac{9}{5}\left({J\over \Delta}\right)^{5} \right]
\ee
and, for~~ $s \ge2$,
\be
D_{s} = (A_2 s +B_2) \left({J\over \Delta}\right)^{s}
\ee
with 
\bea
&&A_2 =  \frac{1}{30(J/\Delta)}\left[30 -110 \left({J\over
\Delta}\right) + 135   
\left({J\over \Delta}\right)^{2} - 54 \left({J\over \Delta}\right)^{3}
\right] \nonumber \\ 
&&B_2 = \frac{1}{15(1-J/\Delta)} 
 \left[5 - 10 \left({J\over \Delta}\right) + 4 \left({J\over
 \Delta}\right)^{2} \right] \nonumber 
\eea


\section{}
\label{appendixB}
\noi For $z=3$, the self-consistent equation. for $P^\star(h)$ [
Eq.~\ref{a.2} ] is a quadratic equation,
\be
[(p_{2} -p_{1})-(p_1 -p_{0})] P^\star(h)^2 + [ 2 (p_{1}-p_{0}) -1]
P^\star (h) + p_{0} = 0.
\label{z3.1}
\ee
For the rectangular distribution, 
the coefficient of ${P^\star}^2$ is zero for a range of $h$-values, and   
$P^\star(h)$ is still a piecewise linear function of $h$
\be
P^\star(h) = \cases{ 0 & {for $\epsilon <  0$ } \cr
       \frac{\epsilon}{1 - 2(J/\Delta)} &
                 {for $0 \le \epsilon \le 1 - 2(J/\Delta)$ } \cr
        1 & {for  $\epsilon > 1- 2(J/\Delta)$ }
}
\label{z3.2}
\ee
where $\epsilon$ is defined as ,
\be
\epsilon = \frac{1}{2} ( 1 + \frac{h}{\Delta} - \frac{3 J}{\Delta} )
\label{z3.3}
\ee  
The self-consistent equation for $Q(x)$ [ Eq.~\ref{Q(x)} ] becomes,
\be
x(p_1-p_0)\left[Q(x)\right]^2
+\left[2xP^\star(p_2-p_1)-1\right] Q(x)
+ x P^{\star 2} (p_3-p_2)+Q_0=0
\label{z3.4}
\ee
where $Q_0$ is obtained [ Eq.~\ref{Q_0} ] as 
\be
Q_0=(1-p_1)-2(p_2-p_1)P^\star
+\left[(p_2-p_1)-(p_3-p_2)\right]P^{\star 2}
\label{z3.5}
\ee
and the expression~(\ref{G(x|h)}~) for $G(x|h)$  becomes ,
\bea
G(x|h)= x &&\left\{
[(Q(x)]^3p(3J-h)+3[Q(x)]^2P^\star p(J-h)
 \right. \nn \\
 &&\left.+3[Q(x)]P^{\star 2}p(-J-h)+P^{\star 3}p(-3J-h)\right\}
\label{z3.6}
\eea
Now in the region B ,
\bea 
&&(p_3-p_2)=(p_2-p_1)=(p_1-p_0)=J/\Delta, \nonumber \\
&& p(3J-2mJ-h)=\frac{1}{2\Delta} \mbox{ for $m=0$ to $3$ } \nn   \\ 
\mbox{ and~~~~ } && P^\star+Q_o=1-J/\Delta \nonumber
\eea

Solving Eq.~\ref{z3.4} and choosing the root which is well behaved for
$x$ near $0$, we get 
\be
Q(x)=\frac{1 - \sqrt{1-4 (J/\Delta) x (P^\star+Q_0)}}{2 (J/\Delta) x}
- P^\star
\label{z3.7}
\ee
and the expression for the integrated distribution ~(~\ref{z3.6}~) becomes 
\be
G(x|h)=\frac{x}{2\Delta}\left[P^\star+Q(x)\right]^3 
\label{z3.8}
\ee
Expanding $G(x)$ in power 
series of $x$,
we  obtain the Eq.~\ref{Gs.B.z3} of sec.~\ref{z3}.

In the region C, $p_3$ saturates to the value $1$,
$p(-3J-h)$ becomes zero and $(p_3-p_2)$ is no longer independent of $h$. 
Substituting the appropriate expressions, we find that 

\be
Q(x)=\left.  \frac{1 - \sqrt{1-4 (J/\Delta) x [(1-3(J/\Delta)-\epsilon)
+(P^\star+Q_0)]}} {2 (J/\Delta) x} -P^\star \right.  
\label{z3.12}
\ee
and
\be
G(x|h)=\frac{x}{2\Delta}\left\{[P^\star + Q(x)]^3-P^{\star 3}\right\}
\label{z3.13}
\ee

We note that the term inside the radical sign in $Q(x)$, and also in
$G(x|h)$, is a 
simple linear function of $x$. It is
thus straightforward to expand it in powers of $x$ using binomial
expansion. This gives 
us the Eq.~\ref{Gs.C.z3} of sec.~\ref{z3}.



\begin{references}
\bibitem{foot} This step is not really necessary if we are only interested
in determining the magnetization at the site $O$. Skipping this step leads
to considerable simplification of the relaxation process: first the spins
of generation $n$ are examined, then those of $(n-1)$ etc. till we finally
examine the spin at $O$. No spin is checked more than once. The resulting
configuration is not fully relaxed, but it is easy to prove that further
relaxation will not change the state of the spin at $O$. The argument can
be extended to show that the probability that an avalanches starting at
$O$ is of size $s$ also is the same in this partially relaxed state as
in the fully relaxed state.
\bibitem{z6} This would happen if  $P^{\star}(h)$ as a function of $h$
shows a `double s' curve. Then there must be at least $4$ values of
$h$ for which the slope of the curve is infinite. This is possible
only if the equation determining $P^{\star}_{disc}$ (variant of
Eq.~\ref{z4.2})  is at least a quartic, hence only if $z \ge 6$.
\end{references}
\end{document}